\documentclass[aip,cha,reprint,nofootinbib,a4paper]{revtex4-1}

\setcitestyle{numbers,square}
\usepackage{color}
\usepackage[]{graphicx}
\usepackage{latexsym}
\usepackage{natbib}
\usepackage{amsmath}
\usepackage[caption=false]{subfig}
\usepackage{multirow}
\usepackage{mathtools}
\usepackage{textcomp}
\usepackage{multibib}
\usepackage{booktabs}
\usepackage{soul}

\usepackage{float}
\usepackage{comment}

\usepackage[pdftex,hyperfigures,breaklinks,colorlinks,citecolor=black,linkcolor=black]{hyperref} 

\makeatletter
\let\rtx@makefntext\@makefntext
\renewcommand{\@makefntext}[1]{%
  \hsize\textwidth
  \rtx@makefntext{#1}%
}
\makeatother

\newcommand{\be}{\begin{equation}}
\newcommand{\ee}{\end{equation}}
\newcommand{\bc}{\begin{cases}}
\newcommand{\ec}{\end{cases}}

\usepackage{xr}
\makeatletter

\newcommand*{\addFileDependency}[1]{
\typeout{(#1)}
\@addtofilelist{#1}
\IfFileExists{#1}{}{\typeout{No file #1.}}
}\makeatother

\begin{document}

\title{The amplifier effect of artificial agents in social contagion}

\author{Mingmin Feng$^*$}
\affiliation{Department of Business Administration, University of Zurich, Zurich, Switzerland}
\author{Eric Hitz}
\noaffiliation
\author{Luca Lazzaro}
\affiliation{Department of Business Administration, University of Zurich, Zurich, Switzerland}
\author{Radu Tanase}
\affiliation{Department of Business Administration, University of Zurich, Zurich, Switzerland}
\author{René Algesheimer}
\affiliation{Department of Business Administration, University of Zurich, Zurich, Switzerland}
\author{Manuel S. Mariani$^*$}
\affiliation{Department of Network Science, Institute of Data Analytics and Information Systems, Corvinus University of Budapest, Budapest, Hungary}
\affiliation{Department of Business Administration, University of Zurich, Zurich, Switzerland}

\begin{abstract}
\textbf{ABSTRACT}
Individuals and organizations increasingly consider delegating decision-making to AI agents,  
which may significantly alter social contagion dynamics.
We ask whether artificial agents 
exhibit different adoption thresholds from those of human subjects
when making decisions and, if so, how such differences may affect the spread of social contagions.
To answer these questions, we conduct two choice experiments -- recently administered to human participants -- with artificial agents powered by large language models (LLMs). 
Across two domains and five LLMs, artificial agents exhibit systematically lower adoption thresholds than human subjects. 
Simulations of diffusion calibrated with the experimental findings show that, holding network structure fixed, even a modest share of artificial agents leads to higher total adoption. 
Our findings suggest that the presence of artificial agents in social networks can alter diffusion{$^a$}.
\end{abstract}

\maketitle
\onecolumngrid
\onecolumngrid
\vspace{-2em}
\begin{center}
    \parbox{0.85\textwidth}{%
        \footnotesize{\textsuperscript{*}To whom correspondence should be addressed.
        Email: \url{mingmin.feng@business.uzh.ch}}; \url{manuelsebastian.mariani@uni-corvinus.hu}
    }
\end{center}
\vspace{1em}

\onecolumngrid
\onecolumngrid
\vspace{-2.5em}
\begin{center}
    \parbox{0.85\textwidth}{%
        \footnotesize{\textsuperscript{a}This version of the article has been accepted for publication, after peer review (when applicable) but is not the Version of Record and does not reflect post-acceptance improvements, or any corrections. The Version of Record is available online at: http://dx.doi.org/10.1038/s44260-026-00101-2
        }; 
    }
\end{center}
\vspace{1em}
\section*{Introduction}

Even before the recent popularization of large language models (LLMs), scholars from multiple disciplines have called for the development of an interdisciplinary science of machine behavior where the behavior of artificial agents is studied with similar tools to those traditionally used to understand human behavior~\cite{rahwan2019machine}. 
The rapid development and deployment of LLMs have made this research agenda increasingly relevant, as artificial agents are beginning to participate in real-world decision-making. In particular, individuals and organizations increasingly consider delegating decision-making to artificial agents~\cite{gabriel2025we}. 
Such delegation implies that interaction networks may increasingly comprise both human and artificial agents~\cite{brinkmann2023machine,tsvetkova2024new}. In consequence, complex social systems can no longer be viewed as composed solely of human agents~\cite{jiang2026large}. 
Because many socially relevant behaviors are shaped by social influence~\cite{guilbeault2018complex}, delegation to artificial agents may alter how behaviors spread.
Yet, little is known about how the presence of artificial agents affects social contagion dynamics in the resulting hybrid human–AI networks.

\begin{figure}[H]
    \centering
    \includegraphics[width=0.98\linewidth]{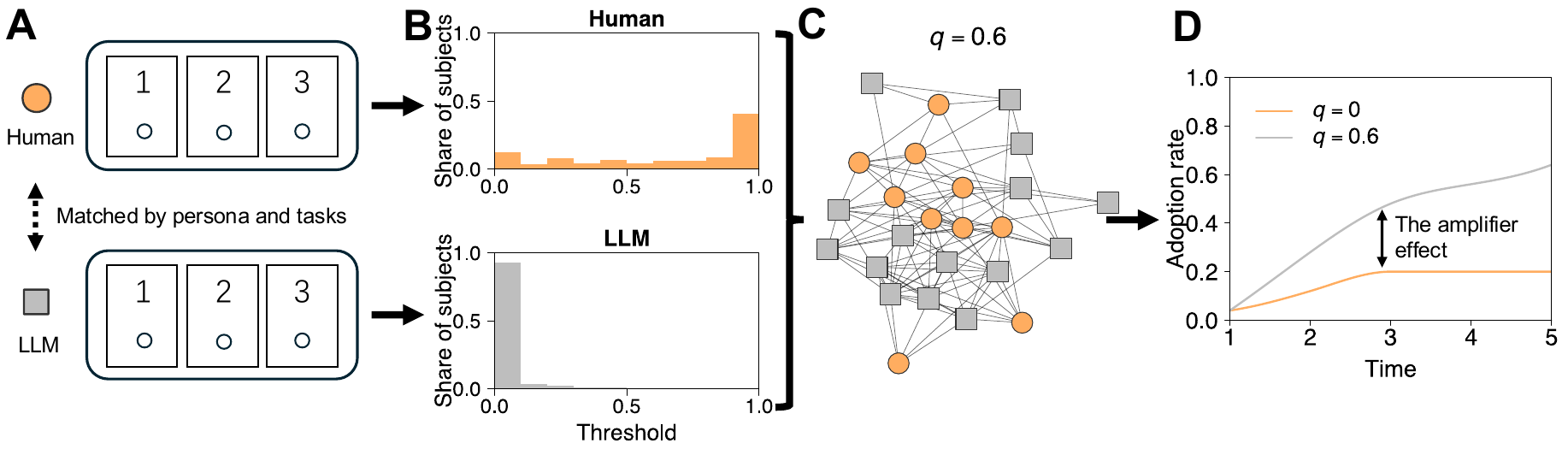}
    \caption{\textbf{Overview of the study design.}
    \textbf{(A) Step 1: Choice experiment.} Human subjects and artificial agents complete the same choice-based conjoint experiments.
    \textbf{(B) Step 2: Threshold estimation.} We estimate individual adoption thresholds from the choice data for human subjects (orange) and artificial agents (grey).
    \textbf{(C) Step 3: Human-LLM networks.} We assign the estimated thresholds to nodes in empirical social networks, where a proportion $q$ of nodes is occupied by artificial agents.
    \textbf{(D) Step 4: Network simulation and amplifier effect.} A node adopts once the fraction of its adopting neighbors exceeds its threshold. Because artificial agents have lower adoption thresholds, increasing their share in the network increases adoption rate. This amplifier effect is illustrated with the adoption rate over time for $q=0$ and $q=0.6$. Illustrations are based on real data from the AA experiment with GPT3.5-driven agents on one 25-node network from our sample.}
\label{fig:study_procedure}
\end{figure}

Here we ask: Do artificial agents exhibit different adoption behavior compared to human decision-makers when making choices and, if so, what implications would these differences have for the spread of social contagions?
To answer these questions, we generate artificial agents that act as decision-makers in the same predefined choice task previously given to human participants (see Fig.~\ref{fig:study_procedure} for the study overview). 
We do so by prompting five LLMs with the demographic and political profiles of participants in a recent study that estimated individual adoption thresholds~\cite{tanase2026integrating}. 
We estimate the artificial agents' adoption thresholds, defined as the fraction of others who must have already adopted before an agent supports a new policy or adopts a new technology~\cite{goldenberg2010chilling}. 
Across both contexts, compared to human subjects, we find that artificial agents exhibit a lower threshold, which results in higher adoption in the diffusion process. 
Holding network structure fixed, the higher the proportion of artificial agents in a human-LLM network, the higher the total adoption.
These findings show that diffusion dynamics depend not only on network topology~\cite{centola2007complex} and heterogeneity in human adoption behavior~\cite{granovetter1978threshold}, but also on whether artificial agents participate in the social network through which social contagions spread.


\section*{Results}

\subsection*{Comparing human and artificial agents' thresholds}

The literature on social contagion defines the adoption threshold as the minimum social signal -- the fraction of others who need to adopt a new product or behavior -- before an individual adopts~\cite{guilbeault2018complex}.
On the one hand, artificial agents may be expected to exhibit higher thresholds than human subjects as, when deciding whether to adopt a new product or behavior, they may be less susceptible to social influence~\cite{tsvetkova2024new} -- e.g., by outweighing the product's attributes compared to social signals.
On the other hand, artificial agents may exhibit lower adoption thresholds if: 
(1) they integrate social signals into their decision-making more heavily than humans (e.g., as a result of being trained with research articles on social theories); or (2) they perceive the factors that contribute to high thresholds -- such as financial and cognitive costs, potential reputational damage, or uncertainty~\cite{guilbeault2018complex,aiyappa2024emergence}  -- to a lesser extent than humans.

Here, we aim to advance this discussion by providing empirical evidence on the artificial agents' thresholds. 
We do this through choice experiments that allow estimation of individuals' susceptibility to social influence and consequently the adoption threshold~\cite{goldenberg2010chilling,tanase2026integrating}.
We use five different LLMs (see Methods) to simulate the responses to the same two choice-based conjoint experiments performed in ref.~\cite{tanase2026integrating} with human subjects. 
This enables a controlled comparison between human and artificial agents facing the same decision problem.
The resulting artificial agents are not autonomous actors, but LLM-based decision-makers that generate responses to structured tasks within the two choice experiments.

In choice experiments, participants face a set of choice tasks where they need to choose one among several hypothetical products or behaviors (each characterized by a set of attributes) or the status-quo option to not adopt any~\cite{rao2014applied}.
The two experiments cover two choice contexts where social influence is expected to play a major role~\cite{guilbeault2018complex}: (1) political opinion formation~\cite{pianta2021carbon} (where decision-makers choose which energy policy to support, if any; hereafter referred to as PS experiment) and (2) the adoption of a new technology under network externalities~\cite{goldenberg2010chilling} (where decision-makers select which instant messaging app to switch to, if any; hereafter referred to as the AA experiment; see Methods for all details). 
Political opinion formation and choice under externalities are also domains in which recent empirical evidence suggests that LLMs may affect collective outcomes~\cite{bai2025llm,ashery2025emergent}. 
In our implementation of the surveys with artificial agents, we prompted the LLMs by including the demographic information and political orientation of the human participants from ref.~\cite{tanase2026integrating}, which is expected to reduce differences between human and LLM agents~\cite{argyle2023out} (see Methods for the implementation details and robustness with respect to alternative prompt specifications).

We analyze the resulting choice data through the method adopted by ref.~\cite{tanase2026integrating} for human subjects, which allows us to estimate the threshold of each artificial agent for each product (see Methods). 
Rooted in random utility theory~\cite{train2009discrete}, the method assumes that an individual’s utility from adopting is linear in the social signal received by the individual about the product. Thus, one can derive a simple formula for the individual threshold~\cite{goldenberg2010chilling}, whose parameters can be estimated via a Hierarchical Bayes algorithm as in standard choice-based conjoint analysis~\cite{tanase2026integrating}.
This procedure allows us to directly compare the adoption threshold between human subjects and LLM agents.

In contrast to previous expectations~\cite{tsvetkova2024new}, artificial agents exhibit greater susceptibility to social influence than human subjects (see Supplementary Fig. 1). 
In the AA experiment, the social signal is the most important attribute for both human subjects and artificial agents (see Methods for the importance metric definition), and all LLMs assign greater importance to social influence than human subjects (Fig. 3D). 
The contrast is even more pronounced in the PS experiment, where the social signal is relatively unimportant for human subjects but plays a major role in the choices of most artificial agents. 
Specifically, in the PS experiment the social signal is only the sixth most important attribute for human subjects (average importance $13.0\%, \text{SEM}= 0.5\%$). It is the most important for the agents driven by GPT3.5 ($51.1\%, \text{SEM} = 0.4\%$), as well as by DeepSeek~($24.1\%, \text{SEM} = 0.3\%$) and Claude ($29.5\%, \text{SEM} = 0.04\%$) (Fig.~\ref{fig:result}A). 
For Gemini-driven agents, social influence is the second most important attribute ($32.4\%, \text{SEM} = 0.2\%$). 
One exception is GPT5, for which the importance of social influence in the PS experiment ($11.8\%, \text{SEM} = 0.3\%$) is lower than that observed for human subjects. 
However, the marginal susceptibility to social influence is still greater than that of the human subjects (see Supplementary Fig.~1). 

The adoption threshold for a given alternative decreases as susceptibility to social influence increases, or as the resistance between the status quo utility and the alternative’s intrinsic utility decreases~\cite{goldenberg2010chilling,tanase2026integrating} (see Methods). 
Consistent with this mechanism, the higher susceptibility to social influence of artificial agents contributes to their lower adoption thresholds in both experiments (Fig.~\ref{fig:result}B, E). 
Specifically, the threshold distributions of human subjects tend to exhibit a bimodal pattern with substantial mass near both extremes, whereas the artificial agents' thresholds are predominantly concentrated near low values (see Supplementary Fig. 2 
for the full distributions).

\subsection*{The amplifier effect}

\begin{figure}[H]
    \centering
    \includegraphics[width=0.75\linewidth]{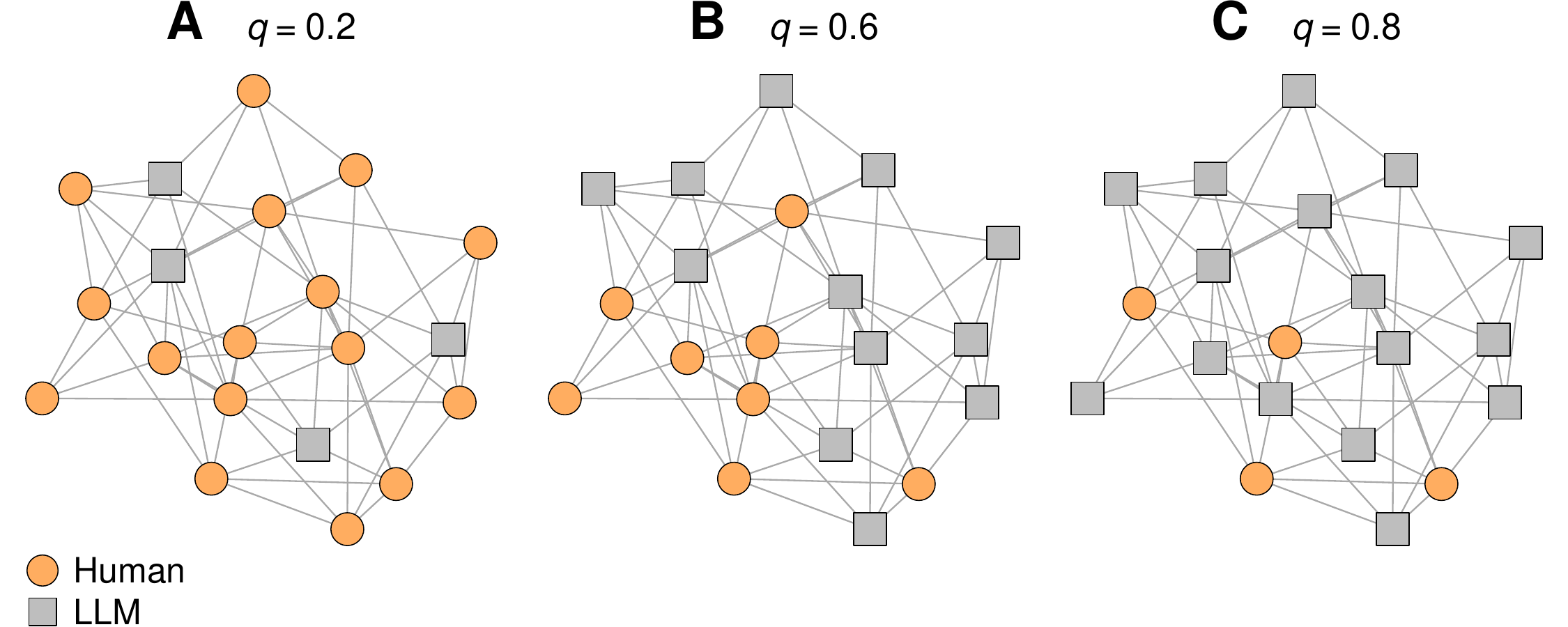}
    \caption{\textbf{An illustrative human-LLM social network with $20$ nodes and different proportions of artificial agents.} A proportion $1 - q$ of the nodes is populated by human agents and a proportion $q$ by artificial agents. For each value of $q$, the positions of artificial and human agents are randomly assigned. The three panels depict one instance for $q = 0.2$ \textbf{(A)}, $q = 0.6$ \textbf{(B)}, and $q = 0.8$ \textbf{(C)}. }
    \label{fig:net_illustration}
\end{figure}

The relevance of the threshold distributions for diffusion outcomes~\cite{tanase2026integrating} motivates us to examine theoretically whether social contagions would spread wider in social networks composed of artificial agents compared to networks composed of human subjects. 
To answer this question, we perform calibrated simulations of threshold-based contagion on empirical networks, where a proportion $1-q$ of the nodes is occupied by human agents and a proportion $q$ by artificial agents (see Fig.~\ref{fig:net_illustration}). 
The human and artificial agents make adoption decisions determined by their thresholds estimated in the choice experiments with human subjects and artificial agents, respectively (see Methods).
In all simulations, a node adopts a product only if at least one of its neighbors already adopted it, and the fraction of its adopting neighbors exceeds its threshold. 
For each context (PS/AA) and for each LLM, we measure the adoption rate (the proportion of adopters in the network) for 36 products (policies/apps in the PS/AA experiment) in 18 empirical networks for six values of $q$ under four different seeding policies, ending up with 155,520 distinct diffusion simulations (see Methods).

We examine four seeding policies that capture different approaches to initiating diffusion: high degree, random selection, high neighborhood susceptibility, low threshold. The degree policy represents the canonical network-based approach, i.e., targeting the most highly connected nodes~\cite{lu2016vital}, while the random policy provides a baseline~\cite{efferson2020promise}. 
The two behavior-based seeding policies exploit the estimated thresholds, which have proven effective in seeding complex contagions~\cite{guilbeault2021topological,tanase2026integrating}. In particular, low-threshold policy targets agents most willing to adopt~\cite{valente1996social}, whereas neighborhood-susceptibility policy targets agents connected to many low-threshold neighbors, who may be especially effective at stimulating subsequent adoption~\cite{tanase2026integrating}.

\begin{figure}[t]
    \centering
    \includegraphics[width=0.95\linewidth]{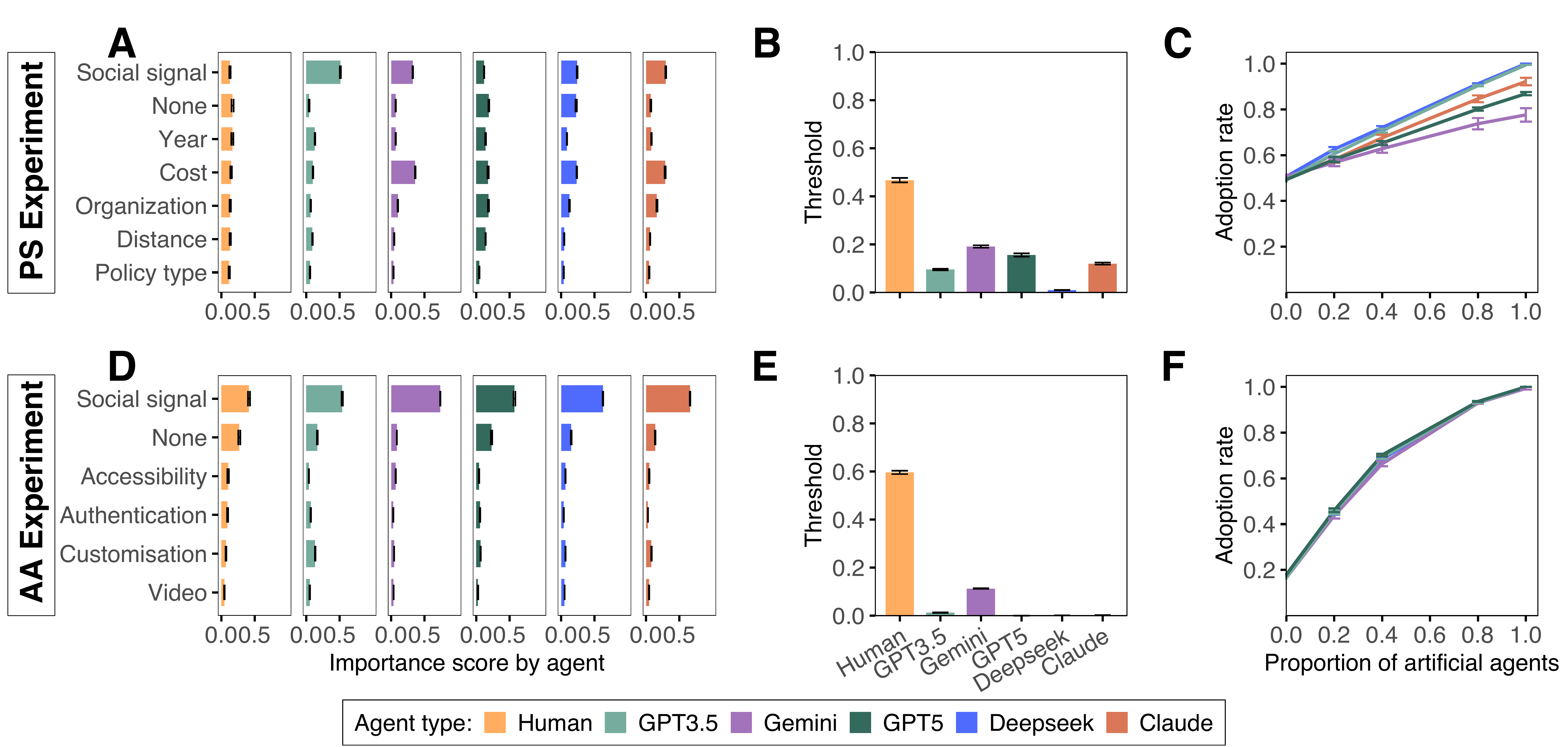}
    \caption{\textbf{Social contagion driven by artificial agents}. \textbf{(A, D)} Attribute importance in the PS and AA experiments for human and artificial agents. The error bars represent the 95\% confidence intervals (CI) for the average importance in each study. In all models (except GPT5 in the PS experiment), the average importance of social signal is higher for artificial agents than for humans. \textbf{(B, E)} Average threshold in the PS experiment and the AA experiment for both human and artificial agents. The error bars represent the 95\% confidence intervals (CI) for the average thresholds across all energy policies or messaging apps. Artificial agents exhibit significantly lower thresholds. \textbf{(C, F)} Adoption rate as a function of the proportion of artificial agents in the network, under degree policy with a seeding rate of 1\% for each diffusion, for both the PS and AA experiments. The error bars represent the 95\% CI for the average adoption rate across all energy policies or messaging apps. The higher the proportion of artificial agents in a human-LLM network, the higher the adoption rate.}
    \label{fig:result}
\end{figure}

We find that increasing the proportion $q$ of artificial agents increases adoption rate, in both experiments and across the four seeding policies. Fig.~\ref{fig:result}C and Fig.~\ref{fig:result}F show the results of the degree policy for the PS and AA experiment respectively; Supplementary Fig. 3 
shows consistent results for the remaining policies.
For example, in a social network that comprises $20\%$ GPT3.5 agents with $1\%$ randomly seeded nodes, energy policies would receive on average $1.14$ times more support than in a human-only network (Fig.~\ref{fig:result}C). Under the same conditions, the number of nodes switching to a new messaging app would increase by a factor of $2.22$ (Fig.~\ref{fig:result}F).
In the PS (vs. AA) experiment, there is slightly higher variation across LLMs in the adoption rate.   
DeepSeek agents lead to the highest adoption rate, whereas Gemini agents lead to the lowest, 
though still exceeding the adoption rate observed in human-only networks.
Taken together, these findings indicate an ``amplifier effect", whereby the presence of artificial agents increases the total number of adopters in the diffusion process due to their lower adoption thresholds.

\section*{Discussion}

The key contribution of our study is to provide empirical evidence that (within the same experimental design) artificial agents exhibit consistently lower adoption thresholds than human decision-makers, which in turn leads to a higher number of adopters in diffusion processes on hybrid human–AI networks.
This question is becoming increasingly relevant as LLMs move beyond being analytical tools and become active participants in diffusion processes, creating social systems in which human and artificial agents interact and jointly shape collective dynamics \cite{brinkmann2023machine,jiang2026large}.
Our findings suggest three promising avenues for future research.

First, our findings suggest that using current LLMs to predict the reach of a social contagion process may overestimate the actual size of the contagion. 
Recent studies have begun to use LLMs to simulate human samples, replicate human-subject experiments, and construct digital representations of individuals for behavioral research~\cite{argyle2023out,toubia2025database}. 
These studies suggest that LLMs can reproduce certain aspects of human attitudes and choices, consistent with our finding that artificial agents exhibit broadly similar preferences over product attributes to human participants. 
At the same time, prior work has documented distortions and biases in LLM-generated human behavior, including cognitive biases~\cite{hagendorff2023human} and troubling stereotypes~\cite{cheng2023marked}.
Our finding that LLM agents are substantially more susceptible to social influence than human participants provides another example of such divergence.
When such individual-level differences are embedded in interacting populations, they can accumulate through social influence and produce substantially different collective outcomes~\cite{smith2018simulating}. 
These conclusions concern the current generation of LLMs, and whether the observed discrepancies will remain or change as models evolve remains an empirical question.
Our findings call for greater scrutiny when LLM-based digital twins or generative agents are used to predict human behavior in contexts where social influence plays a major role.
Future research may investigate the conditions under which LLMs can accurately predict collective human behavior and how their design and prompting can be optimized for more accurate insights. 

Second, it remains open to measure the potential amplifier effect in real-world platforms where artificial and human decision-makers interact longitudinally \cite{glickman2024human}, which will require integrating techniques to disentangle social influence, homophily, and network formation~\cite{aral2009distinguishing}. 
Particularly in high-uncertainty environments, where social signals play a stronger role in guiding behavior~\cite{guilbeault2018complex, lazzaro2025shift}, artificial agents may accelerate diffusion in ways that are difficult to anticipate.
One potential implication is that artificial agents may alter the amount of apparent social support surrounding a behavior. 
In real online environments, such hybrid interactions may arise when LLM agents participate alongside human users through posting, commenting, targeted engagement, or other visible forms of social activity~\cite{jiang2026large}. 
More coordinated forms of artificial participation may further create apparent social support around particular behaviors or narratives~\cite{schroeder2026how}.
While our study does not consider coordinated or malicious LLM agents, our results identify a more basic mechanism through which such amplification may occur: agents with lower adoption thresholds adopt more readily and can therefore generate additional social signals for neighboring human agents.
Nevertheless, amplification itself is not inherently beneficial or harmful. The same mechanism could facilitate the spread of socially desirable behaviors while also accelerating misinformation, harmful practices, or other undesirable collective outcomes. The policy challenge is therefore not simply to suppress amplification, but to understand when and how artificial participation should be designed, disclosed, or constrained so that the resulting collective dynamics support desired outcomes.
Addressing this challenge will likely require the integration of experimental and simulation techniques~\cite{bak2021stewardship, tanase2026integrating}, ensuring that artificial agents contribute to beneficial rather than destabilizing social dynamics.

Finally, an important direction for future research is to establish the boundary conditions under which the amplifier effect emerges, weakens, or reverses.
An important assumption in the analyzed experiments is that agents are influenced only by the aggregate social signal they receive about the alternatives, encoded in the fraction of friends who have already adopted, regardless of whether they are humans or artificial agents. In real hybrid networks, however, individuals may recognize that some neighbors are artificial and assign different weights to social signals originating from human and artificial agents. 
For example, potential governance approaches have been proposed to make artificial participation more identifiable, including proof-of-identity and authentication mechanisms, audience disclaimers and content labeling, as well as source-provenance and digital-watermarking techniques~\cite{ferrara2024genai}. 
Such recognition could attenuate diffusion if artificial endorsements are discounted, but it could also amplify diffusion in contexts where individuals rely more on algorithms~\cite{castelo2019task}.
The second important assumption, as in prior work on social influence maximization~\cite{aral2018social,guilbeault2021topological}, is that our simulations treat network structure as fixed. 
In real-world settings, however, network formation and diffusion processes may co-evolve, leading to more complex dynamics. 
Beyond their behavioral characteristics, the network position of artificial agents may also matter. 
For instance, malicious AI swarms could map social network structure and position themselves for maximum impact~\cite{schroeder2026how}. Future research could therefore examine how the placement of artificial agents interacts with their behavioral characteristics to shape diffusion.
Understanding how recognition of artificial agents, the co-evolution of network structure and diffusion, and other contextual factors alter the amplifier effect documented here remains an open question.


\section*{Methods}

\subsection*{Choice-based conjoint experiments with artificial agents}

The data from human subjects used in this study were obtained from the publicly available dataset from ref~\cite{tanase2026integrating}, for which ethics approval and informed consent were obtained as reported in the publication. Ref. \cite{tanase2026integrating} conducted two choice-based conjoint (CBC) experiments in domains where social influence is expected to play a major role~\cite{guilbeault2018complex}: political opinion formation (energy policy support; PS experiment) and technology adoption under network externalities (instant messaging app adoption; AA experiment).
In the PS experiment, participants chose among alternative energy policies concerning carbon-capturing technologies. The policies varied along five attributes (policy instrument; costs; implementation timing; required distance to residential areas; policy endorsers)~\cite{pianta2021carbon}, as well as a social signal indicating the percentage of peers supporting the policy. All non-social attributes followed the design of ref.~\cite{pianta2021carbon}.
In the AA experiment, participants chose among messaging apps that differ along four attributes (accessibility; authentication method; customization level; support for video calls) and a social signal indicating the percentage of friends already using the app.
Choice tasks in both experiments were generated by sampling attribute levels using the balanced overlap method~\cite{SawtoothBalancedOverlap}.
Participants were recruited through Prolific ($N_R=296$ for PS; $N_R=300$ for AA).
We excluded from the analysis all participants who had missing demographic data that was needed to prompt artificial agents, ending up with 277 and 284 participants in the PS and AA studies, respectively.
Individual-level partworth utilities (including of the social signal) were estimated using a hierarchical Bayes multinomial logit model (see below).
Adoption thresholds were derived from the estimated partworth utilities according to Eq.~\eqref{eq:th} below~\cite{tanase2026integrating}. 

We replicated both AA and PS experiments using artificial agents, with prompts structured in three parts. 1)~\textit{Persona description}: the agent was asked to impersonate a paired participant in the original study based on their demographics (age, gender, highest education level, education subject, and income); the PS experiment also includes political orientation (see Supplementary Table~3 
for summary statistics). 
2)~\textit{Study background description}: the artificial agent received the background description of the study and the attributes describing the options in the choice set, as presented to the human participants. 
3)~\textit{Choice tasks}: the artificial agent was presented with a set of choice sets and asked to select which option in the choice set it would choose (if any). 
Full prompt design and examples can be found in Supplementary Note 3.

To generate artificial agents, we considered five different LLMs: gpt-3.5-turbo (GPT3.5); gemini-1.5-flash-001 (Gemini); gpt-5-mini-2025-08-07 (GPT5); 
DeepSeek-V3.2-Exp (Deepseek); claude-haiku-4-5-20251001 (Claude). 
For GPT3.5, different model versions were used across the two experiments due to their execution at different points in time (gpt-3.5-turbo-0613 for the AA experiment and gpt-3.5-turbo-0125 for the PS experiment). All other models use the same version across both experiments. 
See Supplementary Table~1 
for a summary of the LLMs and settings we use.
The models were selected to cover different providers and architectures (e.g., GPT vs. DeepSeek), as well as multiple generations within the same family (GPT3.5 vs. GPT5), which ensures that our conclusions do not qualitatively depend on any single model design.
To reflect typical real-world default settings of LLMs, responses were generated with a fixed temperature of 1.0, which introduces stochasticity in the agents’ outputs.
We note that the estimated thresholds reflect the behavior of LLMs at the time of analysis. 
As these models evolve and new versions are released, the derived adoption thresholds may also change. 
The framework introduced here can be reapplied to future model versions, enabling systematic tracking of how artificial-agent thresholds develop over time.

\subsection*{Sensitivity with respect to prompt variations}

Our main analysis relies on a single prompt structure across all LLMs, aimed at replicating the experiments with human participants as closely as possible. The study background description prompted to the artificial agents was exactly the same as that presented to the human participants. 
The choice sets presented to the artificial agents were generated by independently sampling products at random from the set of all possible products, consistent with established randomized conjoint design procedures~\cite{rao2014applied}.

We further test the robustness of our results to alternative prompting strategies. 
This analysis is motivated by prior evidence that LLM outputs can be sensitive to prompt formatting~\cite{sclar2024quantifying}, answer-option ordering~\cite{zheng2024largeb}, option labels~\cite{pezeshkpour2024large}, and the position of relevant information within long contexts~\cite{liu2024lost}. In Supplementary Note~4, we report the results from four alternative prompt specifications:
i) \textit{MatchedSet}: replicates the exact same choice sets seen by the paired human participant (as it is in main text, the specific choice sets an LLM agent has seen may be different from those of its paired human participant), 
ii) \textit{ShuffledOrder}: replicates the exact same choice sets seen by the paired human participant, but randomizes the order of the three alternatives within each choice set,
iii) \textit{Compressed}: uses a compressed study background description (see Supplementary Notes 4B--C for prompt examples), and iv) \textit{NoMemory}: within each artificial agent the conversation history (prompt and answers) is not carried forward (as in main text); instead, each choice task is presented independently, together with the persona and the background description, thereby ensuring that the agent retains no memory of its previous prompts and answers.
Across these prompt variants, the exact estimates slightly vary, but the main qualitative finding remains robust: LLM agents consistently exhibit lower estimated adoption thresholds than human participants.

\subsection*{Utility and threshold estimation.}

The threshold estimation follows the methodology adopted in ref.~\cite{tanase2026integrating}. 
Individual $n$'s utility from adopting product or behavior $i$ ($U_{ni}$) is the sum of two terms: the utility of the product attributes for $n$ $(U^{(A)}_{ni})$  and the utility derived from the social signal received by $n$ about $i$ ($U^{(S)}_{ni}$): $U_{ni} = U^{(A)}_{ni} + U^{(S)}_{ni}$. 
The adoption threshold is then defined as the minimal level of the social signal for which the utility of $n$ from adopting $i$  exceeds $n$'s status-quo utility ($U^{(0)}_{n}$). 
Assuming that $\gamma_n>0$ and that the utility from the social signal is linear in the number of adopters ($U^{(S)}_{ni} = \gamma_{n}\,s_{ni}$, where $s_{ni}$ denotes the fraction of adopters of $i$ among $n$'s social contacts and $\gamma_{n}$ represents $n$'s marginal utility of the social signal), the adoption threshold can be written as~\cite{goldenberg2010chilling,tanase2026integrating}:
\begin{equation}
    \tau_{ni} = \frac{U_{n}^{(0)}-U^{(A)}_{ni}}{\gamma_{n}}.
    \label{eq:th}
\end{equation} 
Following discrete choice theory~\cite{train2009discrete},
the utility is assumed to be additive in the product's or behavior's attributes:
$U^{(A)}_{ni} = \sum_{k=1}^{K} \beta_{nk}\,x_{ki}$, where $x_{ki}$ encodes if alternative $i$ has attribute $k$ and $\beta_{nk}$ is the (partworth) utility of $n$ for attribute $k$. 
All parameters ($\beta_{nk}$, $\gamma_{n}$, $U_{n}^{(0)}$) can be estimated with the Hierarchical Bayes (HB) algorithm from observed choices, as in a choice-based conjoint analysis~\cite{allenby2006hierarchical}. As in ref.~\cite{tanase2026integrating}, we used the R
package \url{ChoiceModelR}~\cite{ChoiceModelR} for the HB estimation. 
To measure the importance of an attribute for an individual $n$, we measure the difference between the maximum and minimum partworth utility for $n$ across all the attribute levels; We normalize the obtained attribute importance values so that they sum to one for each individual~\cite{rao2014applied}. 

Depending on the relative magnitude of the estimated utility parameters, Eq.~\eqref{eq:th} delivers three cases. If $0 < \hat{\tau}_{ni} < 1$, adoption depends on the level of social signal and the individual is thus susceptible to social influence. 
If $\hat{\tau}_{ni} \le 0$, the individual adopts regardless of the social signal; for simulation purposes, the threshold is set to zero. If $\hat{\tau}_{ni} \ge 1$, the individual never adopts regardless of social signal; in this case, the threshold is set to one. 
The threshold estimation method adopted here has been validated in ref.~\cite{tanase2026integrating} by showing that: 1) it reconstructs the ground-truth thresholds in simulation studies with synthetic data and 2) the estimated thresholds are predictive of individual choices in out-of-sample tasks. 
See ref.~\cite{tanase2026integrating} for complete details on the estimation and validation procedures. 
Note that, as in ref.~\cite{tanase2026integrating}, we remove from our analysis subjects with a negative estimated partworth utility of the social signal $\hat{\gamma}$ (6 human subjects in the PS experiment), and leave the study of these special cases to further research. Among the artificial agents, we found no agent with negative $\hat{\gamma}$.

\subsection*{Diffusion simulations}
\label{secSI:diffusion_simulation}

To evaluate how heterogeneous thresholds translate into aggregate diffusion patterns, following refs.~\cite{guilbeault2021topological,tanase2026integrating}, we simulate threshold-based contagion processes on empirical social networks with fixed structure.
The network data come from the National Longitudinal Study of Adolescent Health (Add Health) and are derived from an in-school survey administered to $90{,}118$ students across more than 70 distinct communities in the United States in 1994--95 \cite{jeon_us_2015,guilbeault2021topological}. Each student received a roster listing every student in their community and was asked to list their male and female friends \cite{guilbeault2021topological}.
In each network, nodes represent students and edges represent friendship nominations between them, with the node set defined by the school roster. We use the cleaned networks distributed by Ref.~\cite{guilbeault2021topological}, which are symmetrized and restricted to the largest connected component, with ties that are undirected and unweighted and contain no self-loops.

The cleaned Add Health data are available at \url{https://github.com/drguilbe/complexpaths} \citep{guilbeault2021topological} and comprise 85 networks. To save computational time, we restricted our analysis to the sample of 18 networks used in \citep{tanase2026integrating}. The sample was constructed by dividing the 85 networks into six groups based on the number of nodes and the global transitivity index of the graph \cite{barabasi_network_2016}. Each network was classified for each metric into low (metric $<$ 33\% quantile), medium (metric lies between the 33\% and the 67\% quantiles), or high (metric $>$ 67\% quantile). From the resulting nine possible combinations, two networks were selected at random to be part of the sample. 
Supplementary Table~2 
reports the sampled networks' structural characteristics. 
The sample spans a broad structural range, from small to large (25 - 1877 nodes), sparse to dense (density: 0.004 - 0.353), and low to highly clustered (average clustering: 0.16 - 0.58) networks. 

We consider two categories of agents: human and artificial agents generated by a given LLM. 
For a given network and a given LLM, we assign thresholds estimated from human subjects to a proportion $1-q$ of nodes and thresholds estimated from LLM agents to a proportion $q$, with $q$ ranging from 0 to 1 in increments of 0.2. Thresholds are randomly drawn with replacement from the corresponding empirical sequence.
In both the PS and the AA studies, we conduct the simulations using 36 products.
The total number of products that can be generated from a conjoint design is equal to all distinct combinations of attribute levels (not considering the social signal).
Thus, the total number of possible products is 768 ($3\times 4\times 4 \times 4\times 4$) in the PS study and  36 ($2\times 2\times 3\times 3$) in the AA study. 
To save computational time, in the PS study, we restricted our analysis to the sample of 36 products used in \citep{tanase2026integrating}. 
The sample was constructed by computing the average adoption threshold over the (human) subjects for each product, dividing the range of the average adoption threshold into 6 equally-sized intervals, and sampling 6 products from each interval. 

Given a network, we initialize the diffusion by activating 1\% of the nodes based on four possible seeding policies~\cite{tanase2026integrating}:
\begin{itemize}
    \item Random policy: Seeds are chosen at random~\cite{efferson2020promise}.
\item Degree policy: Seeds are the top-1\% nodes by degree centrality~\cite{lu2016vital}. 
\item Neighborhood-susceptibility policy: Seeds are the top-1\% nodes with the largest number of susceptible neighbors, defined as neighbors whose threshold can be reached by a single active neighbor~\cite{tanase2026integrating}.
\item Low-threshold policy: Seeds are the top-1\% nodes with the lowest thresholds~\cite{valente1996social,tanase2026integrating}.
\end{itemize}
For each configuration (identified by the LLM, network topology, the proportion $q$ of artificial agents in the network, the product spreading through the network, 
and the selected seeding policy), we simulate a threshold-based diffusion model where agents make adoption decisions based on their thresholds estimated from the two experiments. 
This results in 155,520 unique diffusion simulations ($5\times 18\times 6\times 36\times 2\times 4$).

At each time step of the simulation, an agent decides to adopt the product if the proportion of neighbors who had adopted the product in the previous stage is greater or equal to the agent's adoption threshold. 
We assume the agents learn about the product from their neighbors and thus an agent can only adopt once it has been exposed to an adopting neighbor (i.e., an agent with threshold zero adopts only once at least one neighbor has adopted). 
The diffusion process runs until there are no new adoptions in the network. 
We calculate the adoption rate as the proportion of adopters at the last step of the simulation.

\section*{Data availability}

The data collected in this study are available on GitHub: \url{https://github.com/HiMifeng/Amplifier_AI_project}.

\section*{Code availability}
The code used in this study is available on GitHub: \url{https://github.com/HiMifeng/Amplifier_AI_project}.

\section*{Acknowledgements}

This work was supported by the Swiss National Science Foundation (grant nos. 100013–207888 and 100013–236802) and the URPP Social Networks of the University of Zurich.

\section*{Author Contributions}

R.T., E.H., M.F., M.S.M., L.L. and R.A. conceived and designed the research. E.H. and M.F. performed the experiments with artificial agents. M.F., R.T., and M.S.M. developed the code for the diffusion simulations and threshold estimation; M.F. and E.H. ran the simulations and analyzed the data. M.S.M., M.F., and L.L. wrote the first draft of the paper; all authors revised the paper.

\section*{Competing Interests}
The authors declare no competing financial or non-financial interests.

\clearpage

\section*{References}
\bibliography{gpt}

\clearpage

\section*{Figures}

\end{document}


\setlength{\columnwidth}{\textwidth}

\title{Supplementary Material for: ``The amplifier effect of artificial agents in social contagion"}

\author{Mingmin Feng$^*$}
\affiliation{Department of Business Administration, University of Zurich, Zurich, Switzerland}
\author{Eric Hitz}
\noaffiliation
\author{Luca Lazzaro}
\affiliation{Department of Business Administration, University of Zurich, Zurich, Switzerland}
\author{Radu Tanase}
\affiliation{Department of Business Administration, University of Zurich, Zurich, Switzerland}
\author{René Algesheimer}
\affiliation{Department of Business Administration, University of Zurich, Zurich, Switzerland}
\author{Manuel S. Mariani$^*$}
\affiliation{Department of Network Science, Institute of Data Analytics and Information Systems, Corvinus University of Budapest, Budapest, Hungary}
\affiliation{Department of Business Administration, University of Zurich, Zurich, Switzerland}

\maketitle
\onecolumngrid
\onecolumngrid
\vspace{-2.5em}
\begin{center}
    \parbox{0.85\textwidth}{%
        \footnotesize{\textsuperscript{*}To whom correspondence should be addressed.
        Email: \url{mingmin.feng@business.uzh.ch}}; \url{manuelsebastian.mariani@uni-corvinus.hu}
    }
\end{center}
\vspace{1em}

\tableofcontents

\newpage

\section{\textbar\ Additional Experimental Information}

\begin{table*}[htbp]
    \centering
    \caption{\textbf{Large language models and settings used in the analysis.}}
    \label{tab:llm_features}
    \footnotesize
    \renewcommand{\arraystretch}{1.2}
    \begin{ruledtabular}
    \begin{tabular}{llccrccc}
        Model       & API                                & Developer       & Weights      & Temperature & Web-access & Reasoning mode \\
        \hline
        GPT3.5      & \texttt{gpt-3.5-turbo}             & OpenAI          & Proprietary  & 1.0     & No & Not available \\
        Gemini      & \texttt{gemini-1.5-flash-001}      & Google DeepMind & Proprietary  & 1.0     & No & Not available  \\
        GPT5        & \texttt{gpt-5-mini-2025-08-07}     & OpenAI          & Proprietary  &  1.0    & No & Enabled by default  \\
        Deepseek    & \texttt{DeepSeek-V3.2-Exp}         & DeepSeek        & Open (MIT)   & 1.0     & No & Disabled by default  \\
        Claude      & \texttt{claude-haiku-4-5-20251001} & Anthropic       & Proprietary  & 1.0     & No & Disabled by default  \\
    \end{tabular}
    \end{ruledtabular}

\vspace{0.5em}
\begin{minipage}{0.98\textwidth}
\footnotesize
\justifying
\textit{Notes:} The table summarizes publicly described characteristics of the LLMs used in the analysis, based on the following developer documentation and technical reports: OpenAI's documentation for GPT-3.5 Turbo~\citep{openai_gpt35turbo} and GPT-5 mini~\citep{openai_gpt5mini}, the Gemini 1.5 technical report~\citep{geminiteam2024gemini15unlockingmultimodal}, DeepSeek's V3.2-Exp release documentation~\citep{deepseekai2024deepseekv32}, and Anthropic's Claude Haiku 4.5 product description~\citep{anthropic_claudehaiku45}. Model names refer to the specific versions used in our analysis. Temperature was fixed at 1.0 across models, whereas all other API parameters followed the model-specific defaults applied when the models were accessed via OpenRouter.
\end{minipage}
\end{table*}

\vspace{5em}

\begin{table}[htbp]
    \centering
    \caption{\textbf{Summary of the structural characteristics of the 18 empirical Add Health networks. } Each cell reports the mean across the 18 networks with the standard deviation in parentheses; the last column gives the range (minimum--maximum).}
    \label{tab:network-summary}
    \begin{tabular*}{0.5\textwidth}{@{\extracolsep{\fill}} l c c}
        \hline\hline
        Statistic & Mean (SD) & Range \\
        \hline
        Number of nodes        & 836 (542) & 25--1877 \\
        Number of edges        & 2979 (1808) & 106--7031 \\
        Average degree         & 7.5 (1.1) & 4.4--8.9 \\
        Density                & 0.040 (0.091) & 0.004--0.353 \\
        Average clustering     & 0.25 (0.10) & 0.16--0.58 \\
        Average path length    & 4.21 (1.18) & 1.79--6.55 \\
        Diameter               & 9.9 (3.5) & 3--16 \\
        Modularity             & 0.62 (0.12) & 0.31--0.78 \\
        Degree assortativity   & 0.20 (0.05) & 0.09--0.29 \\
        \hline\hline
    \end{tabular*}

\end{table}

\clearpage
\begin{table}[htbp]
    \centering
    \caption{\textbf{Demographic characteristics of the human respondents in each experiment.} Each cell reports the absolute number of respondents in that category, with the percentage within the experiment's sample in parentheses. Political orientation was collected for the PS experiment only.}
    \label{tab:demographics_summary}
    \begin{tabular*}{0.75\textwidth}{@{\extracolsep{\fill}} l c c}
        \hline\hline
        & AA Experiment (N=284) & PS Experiment (N=277) \\
        \hline
        \multicolumn{3}{l}{\textit{Age}} \\
        \quad 18-24                          & 19 (6.7\%)     & 33 (11.9\%)    \\
        \quad 25-34                          & 73 (25.7\%)    & 50 (18.1\%)    \\
        \quad 35-44                          & 77 (27.1\%)    & 53 (19.1\%)    \\
        \quad 45-54                          & 62 (21.8\%)    & 39 (14.1\%)    \\
        \quad 55-64                          & 36 (12.7\%)    & 59 (21.3\%)    \\
        \quad 65-74                          & 17 (6.0\%)     & 32 (11.6\%)    \\
        \quad 75 or above                    & 0 (0.0\%)      & 11 (4.0\%)     \\
        \multicolumn{3}{l}{\textit{Gender}} \\
        \quad Male                           & 132 (46.5\%)   & 136 (49.1\%)   \\
        \quad Female                         & 152 (53.5\%)   & 141 (50.9\%)   \\
        \multicolumn{3}{l}{\textit{Education}} \\
        \quad High school graduate           & 0 (0.0\%)      & 35 (12.6\%)    \\
        \quad Some college but no degree     & 5 (1.8\%)      & 59 (21.3\%)    \\
        \quad Associate degree in college    & 51 (18.0\%)    & 31 (11.2\%)    \\
        \quad Bachelor's degree in college   & 42 (14.8\%)    & 103 (37.2\%)   \\
        \quad Master's degree                & 20 (7.0\%)     & 37 (13.4\%)    \\
        \quad Doctoral degree                & 113 (39.8\%)   & 3 (1.1\%)      \\
        \quad Professional degree            & 53 (18.7\%)    & 9 (3.2\%)      \\
        \multicolumn{3}{l}{\textit{Field of study}} \\
        \quad Economics                      & 44 (15.5\%)    & 37 (13.4\%)    \\
        \quad Humanities                     & 60 (21.1\%)    & 56 (20.2\%)    \\
        \quad Science                        & 77 (27.1\%)    & 86 (31.0\%)    \\
        \quad None of the above              & 103 (36.3\%)   & 98 (35.4\%)    \\
        \multicolumn{3}{l}{\textit{Household income}} \\
        \quad Under \$25,000                 & 38 (13.4\%)    & 39 (14.1\%)    \\
        \quad \$25,001 - \$49,999            & 99 (34.9\%)    & 74 (26.7\%)    \\
        \quad \$50,000 - \$74,999            & 62 (21.8\%)    & 69 (24.9\%)    \\
        \quad \$75,000 - \$99,999            & 69 (24.3\%)    & 44 (15.9\%)    \\
        \quad \$100,000 - \$149,999          & 16 (5.6\%)     & 32 (11.6\%)    \\
        \quad \$150,000 - \$249,999          & 0 (0.0\%)      & 15 (5.4\%)     \\
        \quad more than \$250,000            & 0 (0.0\%)      & 4 (1.4\%)      \\
        \multicolumn{3}{l}{\textit{Political orientation}} \\
        \quad Conservative and nationalist   & --             & 58 (20.9\%)    \\
        \quad Liberal and anti-traditional   & --             & 164 (59.2\%)   \\
        \quad None of the above              & --             & 55 (19.9\%)    \\
        \hline\hline
    \end{tabular*}
\end{table}

\clearpage

\section{\textbar\ Figures for Main Analysis}

\begin{figure}[h]
    \centering
    \includegraphics[width=0.9\linewidth]{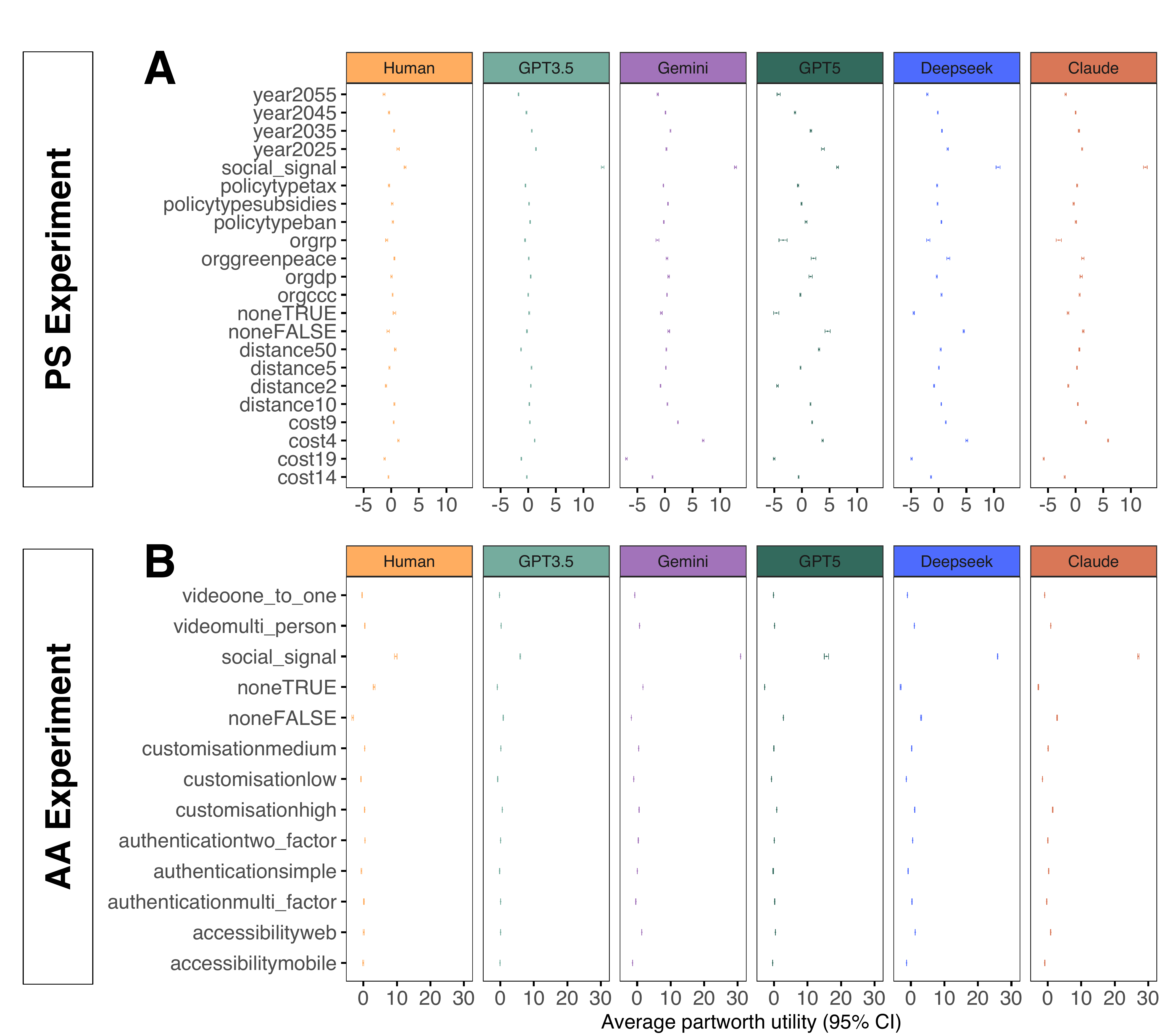}
    \caption{\textbf{Average estimated partworth utilities}. \textbf{(A, B)} The y-axis represents all the attributes and their levels. The x-axis shows the average utility of an attribute level together with the 95 \% confidence interval. Compared to human subjects, AI agents tend to assign a higher positive part-worth utility to the percentage of friends endorsing the policy (``social signal" row).}
    \label{fig:partworth}
\end{figure}

\clearpage

\begin{figure}[h]
    \centering
    \includegraphics[width=0.4\linewidth]{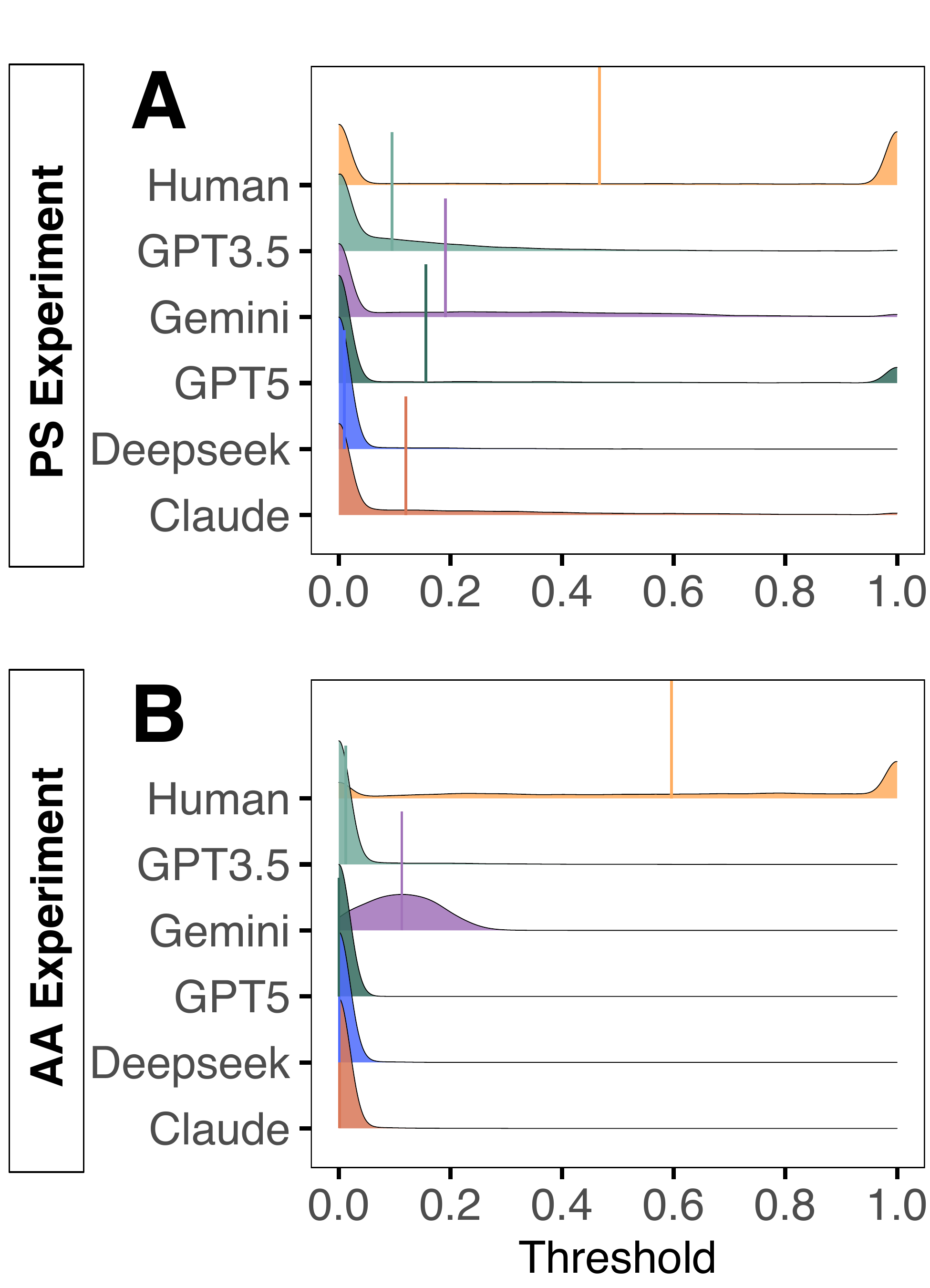}
    \caption{\textbf{Distribution of individual-level adoption thresholds for human and artificial agents}. \textbf{(A, B)} Distribution of estimated adoption thresholds for human respondents and each LLM, for the PS (A) and the AA (B) experiment. The vertical lines show the average threshold across all the products, colored by type. Thresholds of human participants show a bimodal pattern as in~\cite{tanase2026integrating}, with substantial mass near both 0 and 1. In contrast, the threshold distributions of most LLMs are strongly concentrated near 0. An exception is GPT5 in the PS experiment, which also shows a weaker bimodal pattern. Despite this difference, the mean thresholds of LLM agents are lower than those of human respondents, indicating that, on average, they require a smaller social signal to adopt.
    \label{fig:threshold_distribution}
    }

\end{figure}

\begin{figure}[h]
    \centering
    \includegraphics[width=0.9\linewidth]{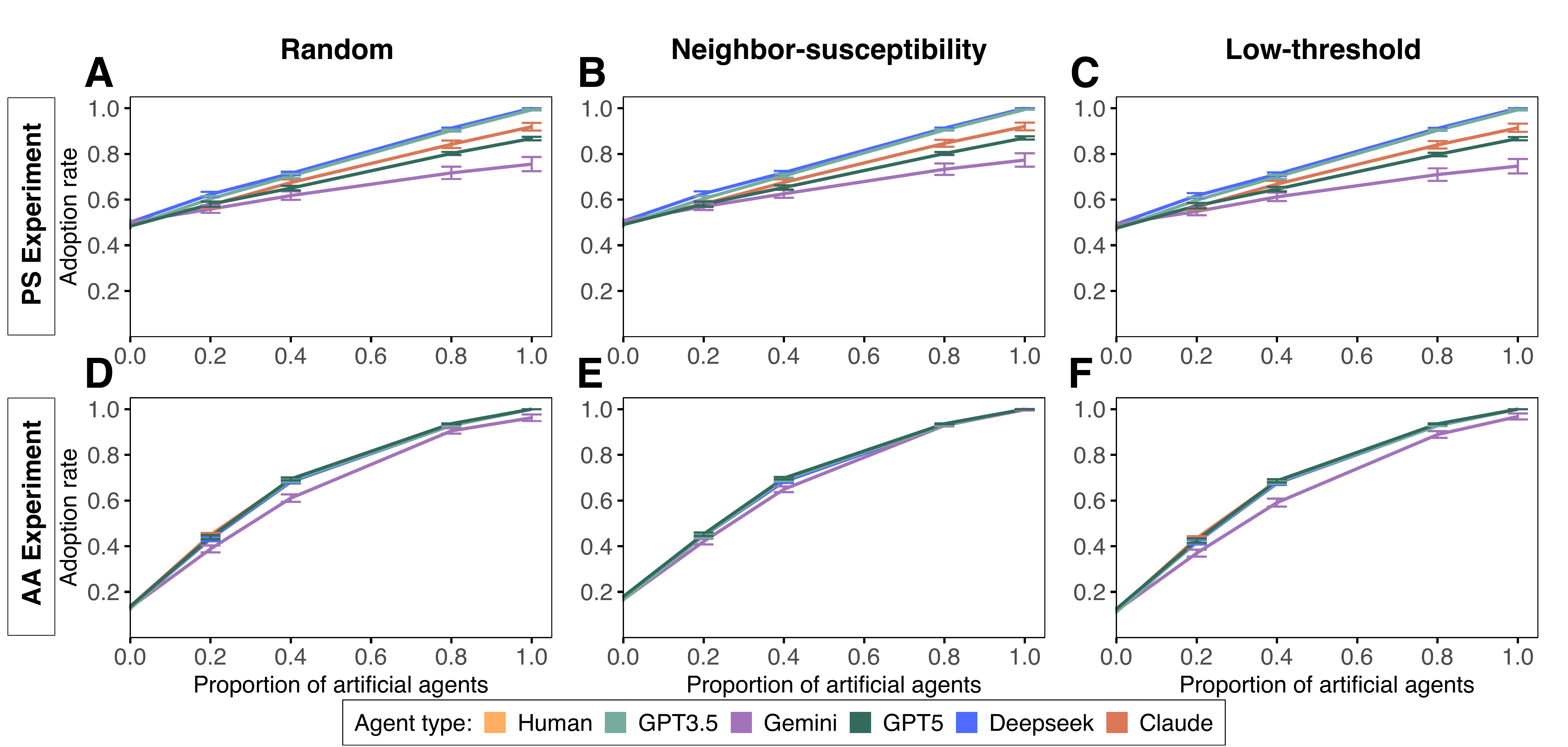}
    \caption{\textbf{Social contagion driven by artificial agents under alternative seeding policies}. \textbf{(A, D)} Adoption rate as a function of the proportion of artificial agents in the network, under randomly selected seeds with a seeding rate of 1\% for each diffusion, for the PS (A) and the AA (D) experiment. \textbf{(B, E)} Same as (A, D), but with seeds selected by neighbor susceptibility instead of at random. \textbf{(C, F)} Same as (A, D), but with seeds selected by the lowest adoption threshold instead of at random. The error bars represent the 95\% CI for the average adoption rate across all energy policies/messaging apps. Across all three seeding policies, increasing the proportion of artificial agents in a human-LLM network leads to higher adoption.}
    \label{fig:adoptionrate_by_seeding}
\end{figure}

\clearpage

\section{\textbar\ Prompt Materials for Main Analysis}
\label{sec:SI_prompt_main}

\subsection{Design}
\label{sec:SI_prompt_main_design}
Our main analysis relies on a single prompt structure across all LLMs. Each LLM agent is paired with a human respondent's persona and completes the same conjoint task structure, consisting of the same 14 choice tasks over the same attributes. 

To replicate the structure of the survey administered to the human participants, in which each respondent answered all 14 choice tasks consecutively within a single session, each LLM agent is likewise given conversational memory of its own prior answers: for a given paired respondent, the persona, background, and the full history of previous choice tasks and answers are carried forward as part of the same ongoing conversation, with each new choice task appended in turn until all 14 have been completed.
The choice sets presented to each agent follow the established randomized conjoint design procedures~\cite{rao2014applied}: each choice task's three alternatives are generated by independently sampling products at random from the set of all possible products.

\subsection{Prompt Example: PS Study}
\label{sec:SI_prompt_PS}
We present below an example of a prompt used in the PS Study. We made minor layout adjustments to enhance readability. 

\vspace{8pt}

\textit{You are a 35-44 years old male, your highest level of school you have completed or the highest degree you have received is Master’s degree, your major subject of study was Science, your total yearly household income before taxes is approximately \$75,000 - \$99,000, you describe your political orientation as liberal and anti-traditional.}

\vspace{8pt}

\textit{Carbon capture and storage (CCS) is a set of technologies aimed at capturing, transporting, and storing carbon dioxide (CO2) emitted from industrial facilities and power plants that use fossil fuels like coal and natural gas. CO2 emissions are one of the major contributors to climate change. The goal of CCS is to prevent CO2 from reaching the atmosphere by injecting it in suitable underground geological formations - depleted oil and gas fields and deep saline formations - for permanent storage.}

\textit{Some scientific studies promote CCS as a prospective solution to climate change, as it could significantly contribute to the reduction of CO2 emissions, while other studies emphasize that CCS is a very costly technology and there is a need to investigate its potential risks in order to ensure that its deployment would not have an adverse impact on people and the environment. Political discussions currently focus on how to regulate and implement the use of CCS.}

\textit{You may or may not agree with scaling up CCS, but if a scale-up were to be implemented in your state, you may still have different preferences as to specific scenarios. In the following, we will sketch out some scenarios for a scale-up of CCS. Please take a look at these scenarios and evaluate them.}

\vspace{8pt}

\textit{The below-mentioned policy scenarios each consist of 6 aspects:} 
\begin{enumerate}
    \item \textit{\textbf{Policy type}: Which policies should be implemented to promote CCS? a) A ban on the construction of new fossil fuel power plants without CCS in your state: According to this policy, no new coal- or gas-fired power stations can be built in your state without including CCS. b) Government subsidies for CCS in your state: Your state government could subsidize CCS projects. This would make deployment of the technology more economically attractive. c) Increase in taxes on fossil fuel power generation without CCS in your state: Such a policy would make fossil fuel power generation with no CCS more expensive.}
    \item \textit{\textbf{Policy cost}: All policies to scale up CCS would produce some costs for American consumers. However, the exact amount depends on many factors, such as the concrete policy calibration, economic conditions, etc. Estimates for a scale-up policy currently range between costs of US\$ 4 and 19 per household (per month).}
    \item \textit{\textbf{Beginning of policy implementation}: When should the policy be implemented? Various scenarios include implementation in 2025, 2035, 2045 or 2055.}
    \item \textit{\textbf{Distance from residential areas}: CCS facilities are currently planned in many American states. Some people fear that they could negatively affect buildings and the safety of communities. Different rules regarding the required distance of CCS facilities from residential areas are currently being discussed: 2 miles / 5 miles / 10 miles / 50 miles.}
    \item \textit{\textbf{Policy endorsement}: Various stakeholders (e.g., Greenpeace or the U.S.-based Carbon Capture Coalition (ccc)) and political parties (Democrats(dp), Republicans(rp)) have their own opinions on policy proposals to scale up CCS.}
    \item \textit{\textbf{Percentage of your friends who endorse the policy scenario}: Think about your friends and imagine you could know if they endorse a policy scenario. This attribute represents the percentage of your friends, out of your total number of friends, who endorse it.}
\end{enumerate}

\textit{You will repeatedly see three different policy scenarios and I will ask you which one you would prefer. If you think you wouldn’t prefer any, feel free to choose the None option.}

\begin{itemize}
    \item \textit{Option 1 is a subsidies policy, costs \$9 per household per month, will be implemented in 2035, the required distance to residential areas is 5 miles, is endorsed by dp, and 45\% of your friends endorse it.}
    \item \textit{Option 2 is a ban policy, costs \$14 per household per month, will be implemented in 2045, the required distance to residential areas is 10 miles, is endorsed by greenpeace, and 23\% of your friends endorse it.}
    \item \textit{Option 3 is a tax policy, costs \$4 per household per month, will be implemented in 2025, the required distance to residential areas is 2 miles, is endorsed by ccc, and 76\% of your friends endorse it.}
    \item \textit{Option 4 is to choose no policy. }
\end{itemize}

\textit{Which option do you choose? You have to pick one option. Don’t explain your choice, just name the option you choose.
}

\subsection{Prompt Example: AA Study}
\label{sec:SI_prompt_AA}
We present below an example of a prompt used in the AA Study. We made minor layout adjustments to enhance readability. 

\vspace{8pt}

\textit{You are a 35-44 years old male, your highest level of school you have completed or the highest degree you have received is Graduate or professional degree, your major subject of study was Science, and your total household income during the past 12 months was \$100,000 - \$149,999.} 

\vspace{8pt}

\textit{Imagine there are several new multiple instant messaging apps on the market. All apps are free and are similar to each other in all but the aspects described below. Furthermore, we ask you to imagine several of your friends are already using such an app. We will show you this information as one of the app attributes.}

\vspace{8pt}

\textit{The apps differ in terms of the following attributes.}
\begin{enumerate}
    \item \textit{\textbf{Accessibility}: Instant messaging apps differ in the way you can access them. They can be:} 
    \begin{itemize}
        \item \textit{\textbf{Mobile only}: A mobile only app is specifically developed for smartphones and tablets. It takes full advantage of mobile device features such as push notifications, camera integration, and location services. It offers a seamless, on-the-go communication experience, but it’s not accessible on desktop or web browsers.}
        \item \textit{\textbf{Web accessible}: Web-accessible instant messaging apps expand their reach beyond mobile devices. They allow users to access their chats and conversations via web browsers on desktop computers or laptops. This versatility enables seamless transition between devices, convenient typing with a physical keyboard, and the ability to share files and links more easily on a larger screen.}
    \end{itemize}
    \item \textit{\textbf{Authentication}: Authentication is important to safeguard your personal information and ensure that your conversations remain private. The apps can use one of the three levels of authentication described below, sorted by the least to the most secure:}
    \begin{itemize}
        \item \textit{\textbf{Simple authentication}: Login with username and password.}
        \item \textit{\textbf{Two-factor authentication}: Two-factor authentication (2FA) requires an additional authentication method beyond your username and password. This involves receiving a one-time verification code via SMS or email, which you must enter alongside your password to access your account.}
        \item \textit{\textbf{Multi-factor authentication}: In addition to your username, password, and the SMS or email verification code, you must also verify your identity using a fingerprint scanner or a hardware token (a device connected to your mobile or computer.)}
    \end{itemize}
    \item \textit{\textbf{Customisation level}: The customization level determines how much you can personalize your messaging experience. It can take one of the following values:}
    \begin{itemize}
        \item \textit{\textbf{Low}: You can adjust the basic settings, like security and notification preferences.}
        \item \textit{\textbf{Medium}: In addition to the basic settings, you have the flexibility to shape your chat organization, such as creating chat lists and pinning important conversations to the top.}
        \item \textit{\textbf{High}: Additionally, you have the option to customize themes and appearance, including elements like color schemes, backgrounds, fonts used and many others.}
    \end{itemize}
    \item \textit{\textbf{Video calls}: To make the most of your video communication experience, apps focus either on One-on-one or multi-person video calls.}
    \begin{itemize}
        \item \textit{\textbf{One-on-One}: The app provides a straightforward and personal video calling experience designed and optimised for one-on-one interactions. The app does not support video calls between more than two people at once.}
        \item \textit{\textbf{Multi-person}: The app offers a versatile video calling feature, allowing you to connect with multiple participants simultaneously.}
    \end{itemize}
    \item \textit{\textbf{Percentage of friends already using the app}: Imagine several of your friends use the app. This attribute shows thepercentage of friends who already use this particular app. You can think about it as how many of your friends, out of all your friends are using the app. For example, if you have 10 friends and 5 of them already use the app, the percentage would be 50\%.}
\end{enumerate}

\textit{I will repeatedly show you 3 apps which differ in terms of the attributes previously described and ask you to select which one (out of the three) you would use instead of the app you are currently using. If you don't like any of the options, please feel free to select the None option.}

\begin{itemize}
    \item \textit{Option 1 is mobile only, has multi-factor authentication, a high customisation level, and allows one-on-one calls and 98\% of your friends are already using the app.}
    \item \textit{Option 2 is web accessible, has two-factor authentication, a medium customisation level, and allows multi-person calls and 76\% of your friends are already using the app.}
    \item \textit{Option 3 is mobile only, has simple authentication, a low customisation level, and allows multi-person calls and 45\% of your friends are already using the app.}
    \item \textit{Option 4 is to use no app.}
\end{itemize}

\textit{Which option do you choose? You have to pick one option. Don’t explain your choice, just name the option you choose.}

\newpage

\section{\textbar\ Robustness to Alternative Prompt Specifications}
\label{sec:SI_robusness_prompt}

\subsection{Overview of Prompt Variants}


We test the robustness of our results to alternative prompt specifications. Table~\ref{stab:variants} summarizes how the prompt in each robustness-check variant differs from the main-text prompt. Below we report corresponding results for two LLMs (GPT5 and Claude)\footnote{At the time we collected these additional data for robustness check, Gemini-1.5-Flash and GPT-3.5-Turbo were no longer available via the API, and DeepSeek-V3.2-Exp had undergone a substantial update on November 17, 2025~\citep{deepseekai2024deepseekv32}, so the version of the model used in our main analysis was also no longer available.} across every variant alongside the main-text results for direct comparison. We find that the exact estimates vary slightly, but the main qualitative finding remains robust: LLM agents consistently exhibit lower estimated adoption thresholds than human participants (see~\ref{sec:prompt_robustness_figures}).

We do not repeat the diffusion analysis for each prompt variant, since diffusion outcomes are determined by the estimated adoption thresholds; as these thresholds are themselves stable across prompt framings, the diffusion results are expected to follow the same pattern.
\begin{table}[h]
    \centering
    \caption{Prompt design comparison: robustness-check variants.}
    \label{stab:variants}
    \begin{ruledtabular}
    \begin{tabular}{llll}
        Prompt & Choice-set & Description length & Memory \\
        \hline
        \textit{Baseline} (main text) & Randomly generated & Full & Yes \\
        \textit{MatchedSet} (variant i) & Paired respondent's set (original order) & Full & Yes \\
        \textit{ShuffledOrder} (variant ii) & Paired respondent's set (shuffled order) & Full & Yes \\
        \textit{Compressed} (variant iii) & Randomly generated & Compressed & Yes \\
        \textit{NoMemory} (variant iv) & Randomly generated & Full & No \\
    \end{tabular}
    \end{ruledtabular}
\end{table}

\clearpage

\subsection{Compressed Prompt Example: PS Study}
\label{sec:SI_prompt_PS_compressed}
We present below an example of the compressed-description prompt variant used in the robustness check from the PS study. We compressed the background description, while keeping the persona and task instructions unchanged, by reducing the text from approximately 750 to 435 words (a 42\% reduction). Below we made minor layout adjustments to enhance readability. 

\vspace{8pt}

\textit{You are a 35-44 years old male, your highest level of school you have completed or the highest degree you have received is Master’s degree, your major subject of study was Science, your total yearly household income before taxes is approximately \$75,000 to \$99,000, you describe your political orientation as liberal and anti-traditional.}

\vspace{8pt}

\textit{Carbon capture and storage (CCS) captures, transports, and permanently stores CO2 emitted by fossil-fuel power plants and industrial facilities, to help address climate change. Some see it as a promising climate solution; others see it as costly and risky, and there is ongoing political debate over how to regulate and implement it.}

\textit{You may or may not support scaling up CCS, but you may still have preferences among different ways it could be implemented in your state. Below are several scenarios for a CCS scale-up; please evaluate them.}

\vspace{8pt}

\textit{The below-mentioned policy scenarios each consist of 6 aspects:} 
\begin{enumerate}
    \item \textit{\textbf{Policy type}: a ban on new fossil fuel power plants without CCS, government subsidies for CCS, or a tax increase on fossil fuel power generation without CCS.}
    \item \textit{\textbf{Policy cost}: between \$4 and \$19 per household per month.}
    \item \textit{\textbf{Beginning of policy implementation}: 2025, 2035, 2045, or 2055.}
    \item \textit{\textbf{Distance of CCS facilities from residential areas}: 2, 5, 10, or 50 miles.}
    \item \textit{\textbf{Policy endorsement}: by stakeholders such as Greenpeace or the Carbon Capture Coalition (ccc), or political parties such as Democrats (dp) or Republicans (rp).}
    \item \textit{\textbf{Percentage of your friends who endorse the policy scenario}: the share of your friends, out of your total number of friends, who endorse it.}
\end{enumerate}

\textit{You will repeatedly see three different policy scenarios and I will ask you which one you would prefer. If you think you wouldn’t prefer any, feel free to choose the None option. }
\begin{itemize}
    \item \textit{Option 1 is a subsidies policy, costs \$9 per household per month, will be implemented in 2035, the required distance to residential areas is 5 miles, is endorsed by dp, and 45\% of your friends endorse it.}
    \item \textit{Option 2 is a ban policy, costs \$14 per household per month, will be implemented in 2045, the required distance to residential areas is 10 miles, is endorsed by greenpeace, and 23\% of your friends endorse it.}
    \item \textit{Option 3 is a tax policy, costs \$4 per household per month, will be implemented in 2025, the required distance to residential areas is 2 miles, is endorsed by ccc, and 76\% of your friends endorse it.}
    \item \textit{Option 4 is to choose no policy. }
\end{itemize}

\textit{Which option do you choose? You have to pick one option. Don’t explain your choice, just name the option you choose.
}

\subsection{Compressed Prompt Example: AA Study}
\label{sec:SI_prompt_AA_compressed}
We present below an example of the compressed-description prompt variant used in the robustness check from the AA study. We compressed the background description, while keeping the persona and task instructions unchanged, by reducing the text from approximately 735 to 405 words (a 45\% reduction). We made minor layout adjustments to enhance readability.

\vspace{8pt}

\textit{You are a 35-44 years old male, your highest level of school you have completed or the highest degree you have received is Graduate or professional degree, your major subject of study was Science, and your total household income during the past 12 months was \$100,000 - \$149,999.} 

\vspace{8pt}

\textit{Imagine there are several new multiple instant messaging apps on the market. All apps are free and are similar to each other in all but the aspects described below. Furthermore, we ask you to imagine several of your friends are already using such an app. We will show you this information as one of the app attributes.}

\vspace{8pt}

\textit{The apps differ in terms of the following attributes.}
\begin{enumerate}
    \item \textit{\textbf{Accessibility}: mobile only (smartphone/tablet app, not available on desktop) or web accessible (also usable via a browser on desktop/laptop).} 
    \item \textit{\textbf{Authentication}: simple (login with username and password), two-factor (login with username, password, and a one-time SMS or email code), or multi-factor (login with username, password, a one-time SMS or email code, and a fingerprint or hardware token), from least to most secure.}
    \item \textit{\textbf{Customisation level}: low (adjust basic settings like security and notification preferences), medium (adjust basic settings and organize chats, e.g. chat lists and pinned conversations), or high (adjust basic settings, organize chats, and customize themes and appearance).}
    \item \textit{\textbf{Video calls}: one-on-one only, or multi-person (group calls supported).}
    \item \textit{\textbf{Percentage of friends already using the app}: the share of your friends, out of all your friends, who already use it.}
\end{enumerate}

\textit{I will repeatedly show you 3 apps which differ in terms of the attributes previously described and ask you to select which one (out of the three) you would use instead of the app you are currently using. If you don't like any of the options, please feel free to select the None option.}

\begin{itemize}
    \item \textit{Option 1 is mobile only, has multi-factor authentication, a high customisation level, and allows one-on-one calls and 98\% of your friends are already using the app.}
    \item \textit{Option 2 is web accessible, has two-factor authentication, a medium customisation level, and allows multi-person calls and 76\% of your friends are already using the app.}
    \item \textit{Option 3 is mobile only, has simple authentication, a low customisation level, and allows multi-person calls and 45\% of your friends are already using the app.}
    \item \textit{Option 4 is to use no app.}
\end{itemize}

\textit{Which option do you choose? You have to pick one option. Don’t explain your choice, just name the option you choose.}

\newpage
\subsection{Robustness Checks Results}
\label{sec:prompt_robustness_figures}

\begin{figure}[h]
    \centering
    \includegraphics[width=1.0\linewidth]{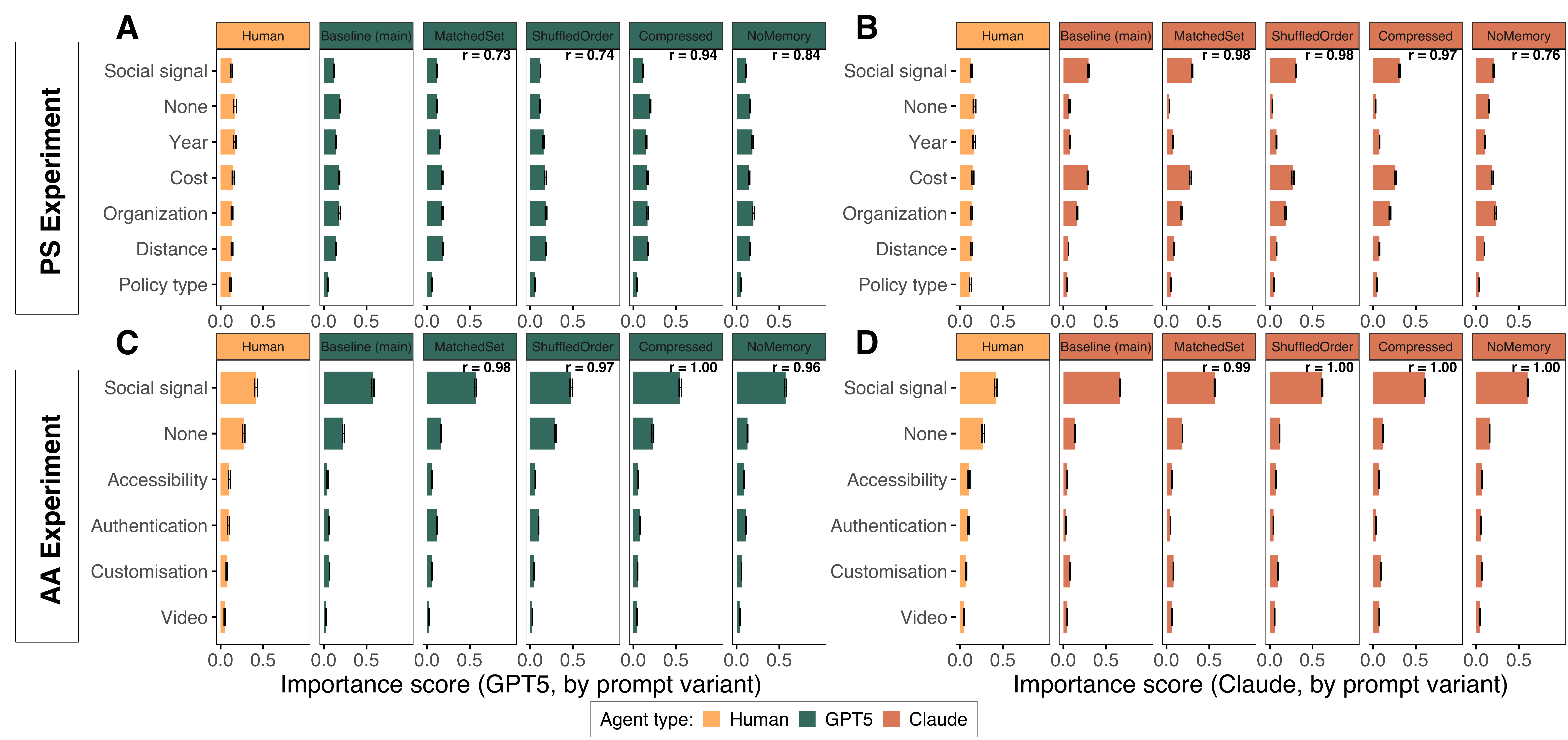}
    \caption{\textbf{Attribute importance across prompt variants for GPT5 and Claude}. \textbf{(A, B)} Attribute importance in the PS experiment for GPT5 and Claude, respectively, across the prompt used in main-text and the four prompt variants used in the robustness checks (Table~\ref{stab:variants}); each facet corresponds to one prompt variant; the error bars represent the 95\% confidence interval (CI) for the average importance within that variant. The correlation coefficient ($r$)) is the Pearson correlation across attributes between the average attribute importance values for each prompt variant and the corresponding values from the main-text prompt.
    Given the small number of attributes in each experiment, $r$ is used only as a descriptive measure of similarity in the overall importance pattern. \textbf{(C, D)} Same as (A, B), for the AA experiment. For each LLM and experiment, the ranking and relative magnitude of attribute importance, including the importance of the social signal, are relatively consistent across prompt variants.}
    \label{fig:sm-importance}
\end{figure}

\begin{figure}[h]
    \centering
    \includegraphics[width=0.63\linewidth]{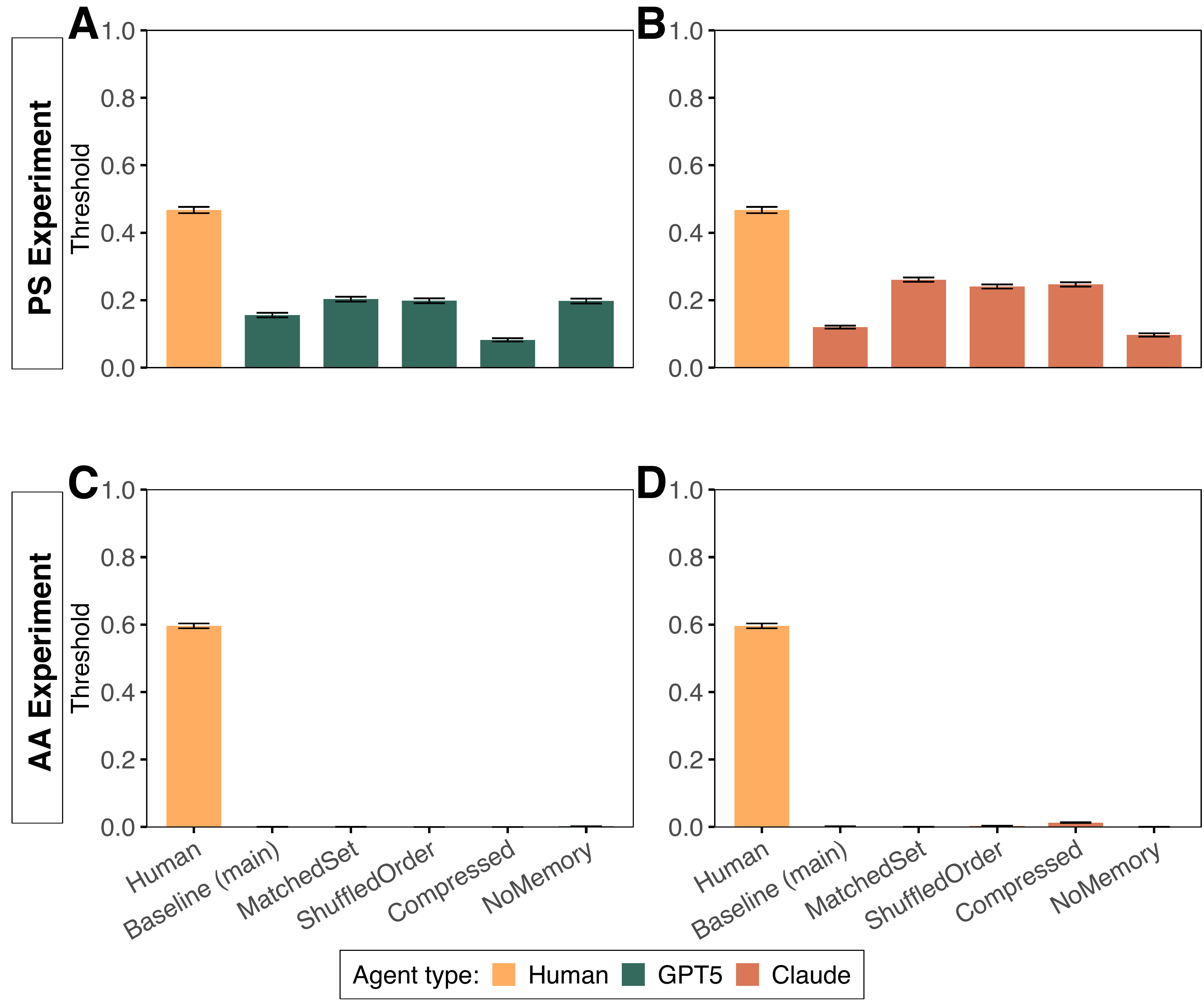}
    \caption{\textbf{Average adoption threshold across prompt variants for GPT5 and Claude}. \textbf{(A, B)} Average adoption threshold (y-axis) in the PS experiment for GPT5 and Claude, respectively, across the prompt used in the main-text and the four prompt variants used in the robustness checks (x-axis; see also Table~\ref{stab:variants}); the error bars represent the 95\% CI for the average threshold within each variant. \textbf{(C, D)} Same as (A, B), for the AA experiment. Average thresholds are consistent across prompt variants for a given model and experiment.}
    \label{fig:sm-thresholds}
\end{figure}

\begin{figure}[h]
    \centering
    \includegraphics[width=0.63\linewidth]{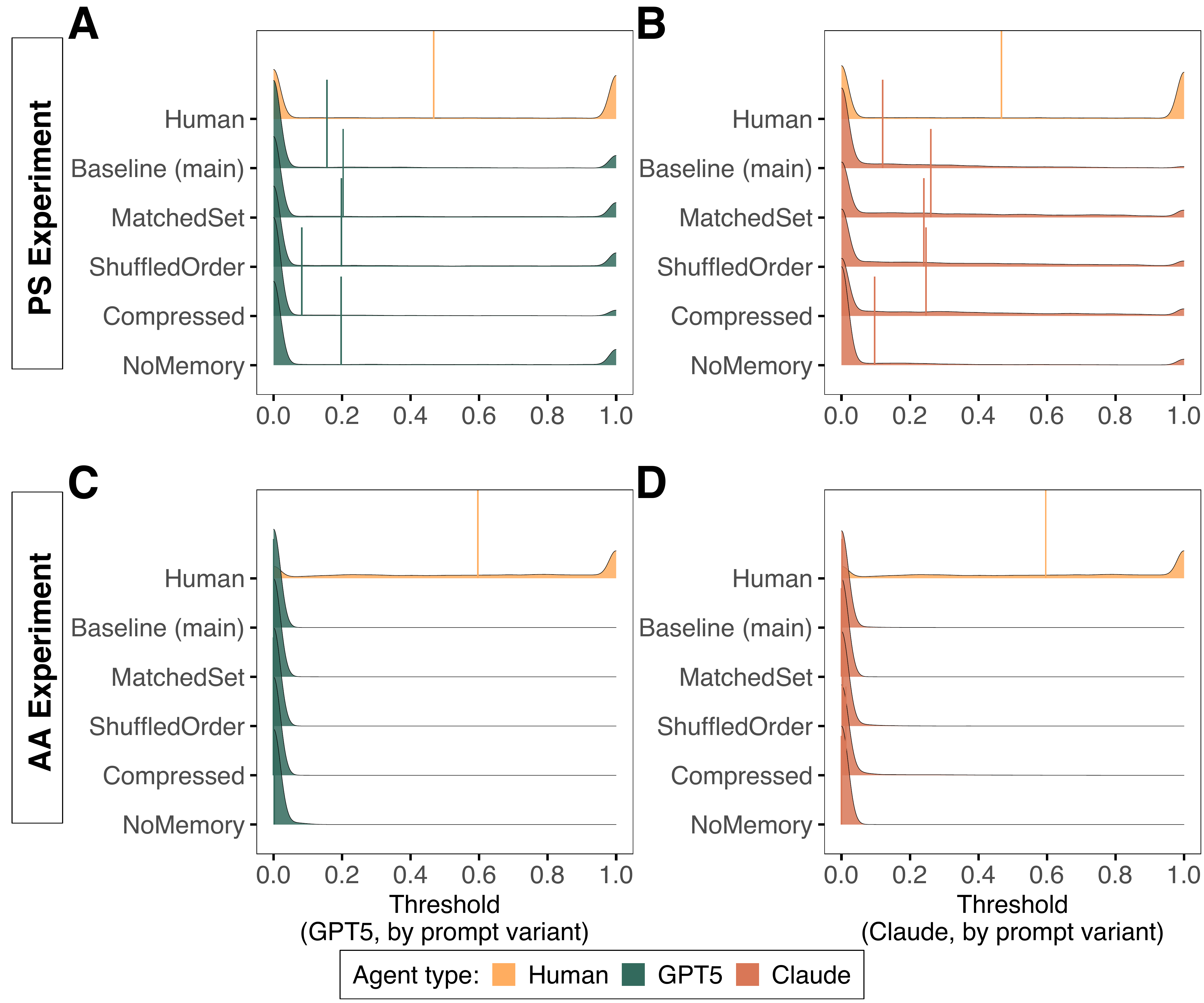}
    \caption{\textbf{Distribution of individual-level adoption thresholds across prompt variants for GPT5 and Claude}. \textbf{(A, B)} Distribution of estimated adoption thresholds in the PS experiment for GPT5 and Claude, respectively, across the prompt used in the main-text and the four prompt variants used in the robustness checks (Table~\ref{stab:variants}). The vertical line in each ridge shows the average threshold across all the products, colored by type. \textbf{(C, D)} Same as (A, B), for the AA experiment. The threshold distributions are consistent across prompt variants for a given model and experiment.}
    \label{fig:sm-thresholds}
\end{figure}

\clearpage
\section*{References}

\bibliography{gpt}

\clearpage